\documentclass[12pt]{article}
\usepackage{amsmath}
\usepackage{amssymb}
\usepackage{graphicx,psfrag,epsf}
\usepackage{enumerate}
\usepackage{natbib}
\usepackage{url} 
\usepackage{color}
\usepackage{bm} 

\begin{document}

\def\spacingset#1{\renewcommand{\baselinestretch}%
{#1}\small\normalsize} \spacingset{1}


  \title{\bf {Bootstrap based uncertainty bands for prediction in functional kriging}
  \author{Maria Franco-Villoria\thanks{Corresponding author, email: {\it maria.francovilloria@unito.it}}\hspace{.2cm}  
    and Rosaria Ignaccolo \\
    Department of Economics and Statistics ``Cognetti de Martiis'',\\
    University of Torino}}
  \maketitle

\bigskip
\begin{abstract}
The increasing interest in spatially correlated functional data has led to the development of appropriate geostatistical techniques that allow to predict a curve at an unmonitored location using a functional kriging with external drift model that takes into account the effect of exogenous variables (either scalar or functional). Nevertheless uncertainty evaluation for functional spatial prediction remains an open issue. 
We propose a semi-parametric bootstrap for spatially correlated functional data that allows to evaluate the uncertainty of a predicted curve, ensuring that the spatial dependence structure is maintained in the bootstrap samples.
The performance of the proposed methodology is assessed via a simulation study.
Moreover, the approach is illustrated on a well known data set of Canadian temperature and on a real data set of PM$_{10}$ concentration in the Piemonte region, Italy. 
Based on the results it can be concluded that the method is computationally feasible and suitable for quantifying the uncertainty around a predicted curve.  
Supplementary material including \texttt{R} code is available upon request.
\end{abstract}

\noindent%
\vfill

\newpage
\spacingset{1.45} 
\section{Introduction}
\label{sec:intro}

Kriging is a well known prediction method in the geostatistics community (see e.g. \cite{geostats}); it allows to predict a (scalar) random field or spatial process $\{Z(s), s \in D\subseteq \mathbb{R}^2\}$ in a new spatial location $s_0$ given a set of observed values $Z = (Z(s_1),\ldots, Z(s_n))$, taking into account the underlying correlation structure.
Spatially dependent functional data (see e.g. the last two chapters of the book by \cite{Kokoszka2012}) have received increasing interest over the last few years. Geostatistical techniques for functional data were first introduced in the pioneering work of \cite{GoulardETal93}, but the development of such techniques is rather recent. The simplest case would be that of ordinary kriging, which allows to predict a curve at an unmonitored site under the assumption of a constant mean (see e.g. \cite{Delicado2010, Giraldo2011,Nerini2010}). The case of a mean function that depends on longitude and latitude was considered in  \cite{Caballero2013,Menafoglio2013,ReyesGM2015}. In their work, \cite{IgnaccoloSERRA} consider more complex forms of non-stationarity, where the mean function may depend on exogenous variables (either scalar or functional), developing the so called kriging with external drift - or regression kriging -  in a functional data setting.
 
 While much effort has been put in prediction, the uncertainty of a predicted curve remains an open issue, since there is no functional version of the kriging variance. The lack of a distribution function in the functional framework leads to the use of resampling methods for confidence band calculation. In this context, \cite{Cuevas2006} consider the standard bootstrap and a smoothed version of it to obtain confidence intervals for location estimators; an informal discussion on the asymptotic validity of the bootstrap approach in a functional framework can also be found in their paper. 
\cite{Goldsmith2013} use a bootstrap approach to account for the uncertainty in Functional Principal Components decomposition in estimanting the functional mean and constructing a confidence band for it.
 Further, \cite{Ferraty2010} propose using ``wild bootstrapping'' in the case of a nonparametric regression model with scalar response and functional covariate and derive asymptotic results; \cite{Rana2016} extend this last work to the case of $\alpha-$mixing dependence.
The recent paper by \cite{Gonzalez2017} shows the consistency of some bootstrap approaches for separable Hilbert-valued random elements but under the assumption of independence. 

While the bootstrap theory is well established for independent data, in the spatial data setting a bootstrap procedure needs to mimic the data generating mechanism in order to reproduce the spatial dependence structure in the bootstrap samples. Mostly by extending bootstrap for time series, several variants of spatial subsampling and spatial block bootstrap methods have been proposed in the literature (see chapter 12 in \cite{Lahiri2003} for a good overview on resampling methods for spatial data). In classical geostatistics, it is common to assume a decomposition of data variability in large- and small-scale components, so that a ``semi-parametric'' bootstrap method as described in \cite{Cressie1993} (p.493) -  and inspired by \cite{Freedman1984} and \cite{Solow1985} - seems appropriate. The latter consists in transforming the residuals of a regression model (i.e. after estimation of the large-scale component) to remove the spatial dependence structure (small-scale) so that resampling can be done on uncorrelated data, to then re-introduce the spatial correlation on bootstrapped samples and finally add up the large-scale component. Recently, \cite{Iranpanah2011} compare the semi-parametric bootstrap with a moving block bootstrap for variance estimation of estimators in a simulation study, and point out some advantages of the semi-parametric approach in terms of precision and accuracy of the estimator. A semi-parametric bootstrap approach has also been considered in \cite{Schelin2010}, where the focus is on the ordinary kriging predictor for data whose distribution is not necessarily Gaussian, and indeed their proposal does not need any distributional assumptions about the data generating process. While \cite{Iranpanah2011} consider the presence of a non-constant mean structure too, the main difference between the two proposals is related to the considered statistics: \cite{Iranpanah2011} suggest to create the bootstrap distribution of the spatial predictor, while \cite{Schelin2010} construct a bootstrap distribution for the contrast defined as the difference between the spatial predictor and the unknown value (the one we want to predict).

The literature available considers either prediction bands for functional data in the case of independent observations or prediction intervals for spatially correlated data but in a scalar framework. In this sense, we fill the gap by providing a solution for spatially correlated functional data. In this framework, however, mimicking the data generating process is expected to be more difficult than in the scalar case. In this paper, we consider the case of functional kriging with external drift (FKED) developed in \cite{IgnaccoloSERRA} and extend it to take into account spatial correlation when estimating the drift functional coefficients by means of an iterative algorithm. To evaluate the uncertainty of a predicted curve, we propose to extend the semi-parametric bootstrap approach for spatially correlated data introduced by \cite{Schelin2010} to the case of functional data, with the addition of a functional drift in the kriging model (a scalar drift was considered in \cite{Iranpanah2011}). Concurrently, \cite{Pigoli2016} have proposed a similar approach to evaluate the kriging prediction error for manifold valued data, but always in terms of a unique value for each location. The extension of the semi-parametric bootstrap to the functional data setting is not straightforward and implies dealing with two main issues: i) the specification and estimation of the spatial dependence structure and ii) the ordering of curves to obtain functional quantiles. In lack of theoretical results about the asymptotic validity of the proposed bootstrap, we rely on a simulation study to evaluate the performance of the proposed method by analysing widths of the functional prediction bands and coverages defined for functional data.

The paper is organized as follows. In Section~\ref{FKED}, we summarize the kriging with external drift methodology whereas in Section~\ref{BOOT} we illustrate the proposed method for deriving prediction bands in the general FKED setting. A simulation study is presented in Section~\ref{SIMU}, followed by an application to two real data sets. All computations are coded in R \citep{R}. A discussion completes the paper.

\section{Functional kriging with external drift (FKED)}
\label{FKED}
Let $\Upsilon_s = \left\{Y_{s}(t); t \in T \right\}$ be a functional random variable observed at location $s \in D \subseteq \mathbb{R}^d$, whose realization is a function of $t \in T$, where $T$ is a compact subset of $\mathbb{R}$. Assume that we observe a sample of curves $\Upsilon_{s_i}$, for $s_i \in D$, $i=1, \ldots, n$, taking values in a  separable Hilbert space of square integrable functions. The set $\left\{\Upsilon_{s}, s \in D \right\}$ constitutes a functional random field or a \emph{spatial functional process} \citep{Delicado2010} that is not necessarily stationary. The following model is assumed:
\begin{equation}
\Upsilon_s = \mu_s + \epsilon_s,
\end{equation}
where $\mu_s$ is the drift describing a spatial trend and $\epsilon_s$  is a zero-mean, second-order stationary and isotropic residual random field, with covariance function $Cov(\epsilon_{s_i}, \epsilon_{s_j}) = C(h), \forall s_i, s_j \in D$ with $h = \left\|s_i- s_j\right\|$, where $h=||s_i-s_j||$ represents the Euclidean distance between locations $s_i$ and $s_j$. At a fixed site $s_i$, $i=1, \ldots, n$, and domain point $t$, the model can be rewritten as a functional concurrent linear model \citep{IgnaccoloSERRA}
\begin{equation}\label{concurrent}
	Y_{s_i}(t)= \mu_{s_i}(t) + \epsilon_{s_i}(t)
\end{equation}
where $\epsilon_{s_i}(t)$ represents the residual spatial functional process $\{\epsilon_{s}(t), t \in T, s \in D \}$ at site $s_i$.
The drift term can be expressed in terms of a set of scalar and functional covariates:
\begin{equation}\label{drift}
\mu_{s_i}(t)= \alpha(t) + \sum_p \gamma_p(t) C_{p,i} + \sum_q \beta_{q}(t)X_{q,i}(t)
\end{equation}
where $\alpha(t)$ is a functional intercept, $C_{p,i}$ is the $p^{th}$ scalar covariate at site $s_i$, $X_{q,i}$ is the $q^{th}$ functional covariate at site $s_i$ and $\gamma_p(t)$ and $\beta_{q}(t)$ are the covariate functional coefficients. Model~(\ref{drift}) parameters can be estimated by means of a generalized additive model (GAM) representation using the R package \texttt{mgcv} (see \cite{IgnaccoloSERRA}, \cite{wood2006} and \cite{mgcvR} for details). The generalized additive model representation of Model (\ref{concurrent}) can be re-expressed as a mixed effects model \citep{Robinson1991, Speed1991} whose parameters are estimated using REstricted Maximum Likelihood (REML) \citep{wood2011} under the assumption of Gaussianity for $\epsilon_{s_i}(t)$ with a longitudinal point of view.

To take into account the spatial correlation between functional observations when estimating the drift term, we propose an iterative algorithm that considers the term $\epsilon_{s_i}(t)$ as a functional random intercept (i.e. a location specific smooth residual) with a given covariance structure, similarly to \cite{FAMM}. An iterative algorithm is also proposed in \cite{Menafoglio2013} to estimate drift coefficients for scalar covariates in universal kriging, as well as in \cite{IgnaccoloCollocation} to take into account the heteroskedasticity of functional residuals.
The algorithm can be summarized as follows:
\begin{enumerate}
\item Fit a standard functional concurrent linear model; estimate the drift term $\mu_{s_i}(t)$ following Model~(\ref{drift}) assuming independent functional observations and obtain the functional residuals $e_{s_i}(t)=Y_{s_i}(t)-\hat{\mu}_{s_i}(t)$ .

\item Estimate the correlation matrix $K=\left\{Corr(\epsilon_{s_i}(t),\epsilon_{s_j}(t))\right\}_{i,j=1,\ldots,n}$ of the residual spatial functional process  using the trace-semivariogram \citep{Giraldo2011}. This is defined, for a zero-mean weakly-stationary isotropic process, as $\upsilon(h)=\int_T \frac{1}{2}Var\left(\epsilon_{s_i}(t)-\epsilon_{s_j}(t)\right)dt$. The trace-semivariogram can be estimated as:
\[
\hat{\upsilon}(h)= \frac{1}{2 \left| N(h)\right|} \sum_{i,j \in N(h)} \int_T \left(e_{s_i}(t) - e_{s_j}(t) \right)^2 dt
\]
where $N(h) = \{(s_i, s_j) : \left\|s_i- s_j\right\| = h\}$. 
The estimate becomes computationally efficient when data are expressed using cubic B-splines, as integration can be avoided by re-expressing the integral in terms of the spline coefficients and basis \citep{Giraldo2011}. Once estimated, the empirical trace-semivariogram provides a cloud of points $(h_g, \hat{\upsilon}(h_g)), g=1,\ldots, G$ to which a parametric model (e.g. exponential, spherical, Mat\'ern) can be fitted as in classical geostatistics using, for example, weighted least squares \citep{Cressie1993}. 

\item Fit Model~(\ref{concurrent}) considering the term $\epsilon_{s_i}(t)$ as a functional random effect, where the inverse of the estimated correlation matrix $\hat{K}$ (dim$(\hat{K})=n\times n$) is used as the precision matrix of a random field across locations, as proposed in \cite{FAMM}. 
In practice this can be implemented using the function \texttt{gamm} in the \texttt{mgcv} package.

\end{enumerate}

The algorithm's convergence is determined based on the Akaike Information Criterion $AIC$, since the effective degrees of freedom may change from iteration to iteration. The algorithm stops when the $AIC$ rate is smaller than 0.1\%, where the criterion rate at the $j^{th}$ iteration is calculated as
\[
	AICrate=\left|\frac{AIC^j-AIC^{j-1}}{AIC^{j-1}}\right|.
\]

The resulting functional residuals (at the last iteration) $e_{s_i}(t)=Y_{s_i}(t)-\hat{\mu}_{s_i}(t)$ can be used to predict the residual curve at an unmonitored site $s_0$ via one of three kriging options: 1) ordinary kriging for functional data \citep{Giraldo2011}, according to which $\hat{e}_{s_0}(t)= \sum_{i=1}^n \lambda_i e_{s_i}(t)$, with kriging coefficients $\lambda_i \in \mathbb{R}$; 2) continuous time-varying kriging \citep{Giraldo2010}, where the kriging coefficients $\lambda_i(t)$ now depend on $t$ and 3) functional kriging total model \citep{Giraldo2009, Nerini2010}, where the kriging coefficients are defined on $T \times T$ and 
$
	\hat{e}_{s_0}(t)= \sum_{i=1}^n \int_T\lambda_i(\tau,t) e_{s_i}(\tau)d\tau.
$
Prediction at the unmonitored site $s_0$ is obtained by adding up, as in the classical regression kriging, the two terms, i.e. 
\[
\hat{Y}_{s_0}(t) = \hat{\mu}_{s_0}(t) +\hat{e}_{s_0}(t), \qquad \text{where} \qquad \hat{\mu}_{s_0}(t) = \hat{\alpha}(t) + \sum_p \hat{\gamma}_p(t) C_{p,0} + \sum_q \hat{\beta}_{q}(t) X_{q,0}(t)
\]
 depends on the covariate values $C_{p,0}$ and $X_{q,0}(\cdot)$ at site $s_0$.

From now on we focus on the ordinary kriging case for the residual field; this contributes to keep a moderate computational complexity in what follows (the two other cases will be considered in the discussion).
Indeed, the ordinary case turns to be computationally convenient because of the trace-variogram's use. Not only integration can be avoided when using B-splines, but also it is possible to show (see appendix) that the trace-variogram induces a covariance structure that is separable with respect to space and the domain of the functional data (e.g. time or depth). Moreover, \cite{Menafoglio2016} show that the solution to the Ordinary Kriging Problem via the trace-covariogram turns out to be the best finite-dimensional approximation of the operatorial kriging predictor for Hilbert-space valued random fields.

Note that, in practice, data are gathered as a finite discrete set of observations $(t_j,y_{ij})$, $t_j \in T$, $j=1,\ldots,M$, $i=1,\ldots,n$. Thus, before fitting Model (\ref{concurrent}), raw data can be transformed into functional observations assuming $y_{ij}=Y_{s_i}(t_j)+ \delta_{ij}$, where $\delta_{ij}$ represents measurement error and $Y_{s_i}(\cdot)$ is a continuous function that
corresponds to a realization of the functional random field $\left\{\Upsilon_{s}, s \in D \right\}$ at site $s_i$. The conversion from discrete data to curves involves smoothing; we use cubic B-splines and the R package \texttt{fda} \citep{fdaR}, choosing the number of basis functions and penalty parameter using functional cross-validation \citep{IgnaccoloSERRA}.

\section{Bootstrap uncertainty bands for functional kriging}
\label{BOOT}

To evaluate the uncertainty of a predicted curve $\hat{Y}_{s_0}(t)$ at a new site $s_0$, we propose a semi-parametic bootstrap approach for spatially correlated functional data that builds on the work done by \cite{Schelin2010} and \cite{Iranpanah2011} for scalar data. The main idea is to decorrelate the data so that resampling can be done on independent observations and then transform back, ensuring that the spatial dependence structure is maintained in the bootstrapped samples.
After removing the drift as in \cite{Iranpanah2011}, we propose to follow \cite{Schelin2010} according to which in the following algorithm a sample of size $n+1$ is drawn and an augmented covariance matrix created, such that a bootstrap datum $Y^{*j}_{s_0}$ is generated at the unmonitored location $s_0$.

Suppose that $\hat{Y}_{s_0}(t) - Y_{s_0}(t)$ follows the distribution $F_n$, a $1-\alpha$ prediction interval for $Y_{s_0}(t)$ can be built as $(\hat{Y}_{s_0}(t)-q_{1-\alpha/2},\hat{Y}_{s_0}(t)-q_{\alpha/2})$, with $q_{\alpha}$  the $\alpha^{th}$ quantile of the unknown distribution $F_n$. The idea is to construct $B$ bootstrap replicates $\{\hat{Y}^{*j}_{s_0},Y^{*j}_{s_0}\}_{j=1}^B$ and approximate $F_n$ by $\hat{F}_n^*$, the empirical distribution of $\{\hat{Y}^{*j}_{s_0}-Y^{*j}_{s_0}\}_{j=1}^B$. The bootstrapping algorithm can be summarized as follows:

\begin{enumerate}
	\item Iteratively estimate the drift following Model~(\ref{drift}) as proposed in steps 1-3 of the FKED algorithm to take into account the spatial correlation and take the functional residuals $e_{s_i}(t) = Y_{s_i}(t) - \hat{\mu}_{s_i}(t)$. 
    \item Estimate the functional residuals covariance matrix $\Sigma$ through the estimated trace-semivariogram. Resampling directly from the functional residuals is not appropriate due to spatial dependence; instead, we can transform them first as it is usual practice when bootstrapping spatial data.
Using Cholesky decomposition, $\hat{\Sigma}_{n\times n}=\hat{L}_{n\times n}\hat{L}_{n\times n}^T$ and the functional residuals can be transformed so that they become (spatially) uncorrelated:
    \[
        \zeta_{n\times M}=\left(\zeta(s_1),\ldots,\zeta(s_n)\right)'=\hat{L}_{n\times n}^{-1}\left(Y_{n \times M}-\hat{\mu}_{n \times M}\right).
    \]
   
     \item Generate $B$ bootstrap samples with size $n+1$, $\zeta^*_{n+1}=\left(\zeta^*(s_1),\ldots,\zeta^*(s_n),\zeta^*(s_{n+1})\right)'$ by sampling with replacement from $\zeta(s_1),\ldots,\zeta(s_n)$.
    \item Create the augmented covariance matrix $\hat{\Lambda}=
    \begin{bmatrix}
    \hat{\Sigma} & \hat{c}_n^T \\
    \hat{c}_n & \hat{\sigma}^2
  	\end{bmatrix}$, 
    where $\hat{c}_n=\{\hat{C}(s_i-s_0)\}_{i=1}^n$, $\hat{C}$ is the estimated covariance function and $\hat{\sigma}^2=\hat{C}(0)$ is the estimated scale. Use Cholesky decomposition so that $\hat{\Lambda}=\hat{R}\hat{R}^T$ and transform the bootstrap samples $\zeta^*_{n+1}$ as
 			\[
 				(e^*(s_1),\ldots,e^*(s_n),e^*(s_0))'=\hat{R}_{(n+1)\times (n+1)}\zeta^*_{(n+1)\times M}.
 			\]
\item The final bootstrap sample is determined as $Y^*_{s_i}(t)=\hat{\mu}_{s_i}(t) + e^*_{s_i}(t)$, $i=1,\ldots,n$ and $Y^*_{s_0}(t)=\hat{\mu}_{s_0}(t) + e^*_{s_0}(t)$.
\end{enumerate}

The bootstrap samples $\{Y^{*j}_{s_1},\ldots,Y^{*j}_{s_n}\}_{j=1}^B$ are then fed into the FKED model to obtain $B$ prediction curves $\hat{Y}^{*j}_{s_0}$ and the differences $\{\hat{Y}^{*j}_{s_0}-Y^{*j}_{s_0}\}_{j=1}^B$ are considered. The prediction interval for $Y_{s_0}(t)$  can be written as 
\[\left(\hat{Y}_{s_0}(t)-q_{1-\alpha/2}^*,\hat{Y}_{s_0}(t)-q_{\alpha/2}^*\right),
\] with $q_{\alpha}^*$ the $\alpha^{th}$ percentile of $\hat{F}^*_n$, that can be obtained ordering the set of curves $\{\hat{Y}^{*j}_{s_0}-Y^{*j}_{s_0}\}_{j=1}^B$. 
However, the idea of ordering curves is not as straightforward as ordering scalar values, and to our knowledge there is no gold standard for doing so. We consider two different ordering techniques available in the literature. 

The first one builds on the idea of band depth \citep{PintadoRomo2009}, that can be defined for any set of $k$ curves.
In their paper, \cite{PintadoRomo2009} suggest using $k$=3, stating that there is no need to increase $k$ as ``the band depth induced order is very stable in $k$''. Even for $k$=2, the computational cost is considerable when the sample includes a large number of functional curves; to improve computation times, we have adopted the fast algorithm proposed in \cite{SunGenton2011} and \cite{SunGenton2012} with $k$=2. 
The band in $\mathbb{R}^2$ delimited by the curves $y_{i_1},y_{i_2}$ is defined as
\[
        B(y_{i_1},y_{i_2})=
        \left\{(t,y(t)):t \in T, min_{r=1,2} \,y_{i_r}(t)\leq y(t) \leq max_{r=1,2}\,y_{i_r}(t)\right\}.
\]  
The sample band depth (\emph{BD}) of a curve $y(t)$ in a set of $n$ curves can be calculated as
\[
        BD_{n,2}(y)=\left(
        \begin{array}{c}
        n\\
        2
        \end{array}
        \right)^{-1}\sum_{1\leq i_1<i_2\leq n} \mathbb{I}\left\{G(y)\subseteq B(y_{i_1},y_{i_2})\right\}
\]
where $\mathbb{I}$ is the indicator function and $G(y)$ is the graph of a curve $y(t)$ defined as the subset of the plane $G(y)=\{(t, y(t)): t \in I\}$. 
Potential problems of using $k=2$ include ties (i.e. more than one curve with the same depth value) and crossing over of the curves delimiting the band (in which case the band
``is degenerated in a point and, with probability one, no other curve will be inside this band''; see \cite{PintadoRomo2009}). To avoid these problems and still count with the computational advantage of using $k=2$, band depth can be modified to take into account whether a portion of the curve is in the band, giving rise to the modified band depth (\emph{MBD}), defined as
        \[
            MBD_{n,2}(y)=\left(
        \begin{array}{c}
        n\\
        2
        \end{array}
        \right)^{-1}\sum_{1\leq i_1<i_2\leq n}\frac{\lambda\left(\left\{t \in T: min_{r=i_1,i_2}y_r(t)\leq y(t)\leq max_{r=i_1,i_2}y_r(t)\right\}\right)}{\lambda\left(T\right)}
        \]
where $\lambda$ is the Lebesgue measure on $T$ (for further details see \cite{PintadoRomo2009}).
With this scheme, the bigger the band depth value, the more central the curve is.

The second ordering scheme is based on $L^2$ distance between curves \citep{Cuevas2006}. In this case, the bootstrap-based predicted curves are ordered based on how distant they are from the zero curve, according to the $L^2$ distance definition:
\begin{equation*}
|| x -y ||= \left(\int_T\left(x(t)-y(t)\right)^2dt\right)^{1/2}.
\end{equation*}
With this scheme, the smaller the distance, the more central the curve is. 

For a confidence level $1-\alpha$, the lower/upper limits of a $100(1-\alpha)\%$ prediction band are obtained by taking either the pointwise (w.r.t. $t$) minimum/maximum of the $100(1-\alpha)\%$ deepest curves using band depth (i.e. those closest to the center of the distribution) \citep{SunGenton2011} or of the $100(1-\alpha)\%$ curves closest to the zero curve using $L^2$ distance \citep{Cuevas2006}.

To evaluate the performance of the bootstrap method proposed, we can use different indicators. First of all, we consider the width of the resulting $100(1-\alpha)\%$ prediction band $\mathcal{B}_{s_0}(\alpha)$ at the unmonitored site $s_0$. Then we evaluate the ``domain coverage'' $DC_{s_0}(\alpha)$, defined as the proportion over the domain $T$ of the simulated curve within the prediction band, calculated as
\begin{equation}\label{DC}
	DC_{s_0}(\alpha)=\frac{1}{M}\sum_{j=1}^M\mathbb{I}\left\{\left(t_j,Y_{s_0}(t_j)\right)\in \mathcal{B}_{s_0}(\alpha)\right\},
\end{equation}
where $M$ is the number of points used for discretizing the curve $Y_{s_0}(t)$. The latter should not be understood as coverage in the classical sense, and thus we propose a ``functional coverage'' $FC_{s_0}(\alpha)$ defined as the percentage of times that the prediction band contains the true curve in a set of $S$ simulations, and calculated as
\[
	FC_{s_0}(\alpha)=\frac{1}{S}\sum_{sim=1}^S\mathbb{I}\left\{(t,Y_{s_0}(t))\in \mathcal{B}^{(sim)}_{s_0}(\alpha)\right\}=\frac{1}{S}\sum_{sim=1}^S\mathbb{I}\left\{DC_{s_0}^{(sim)}(\alpha)=1\right\},
\]
where $\mathcal{B}^{(sim)}_{s_0}(\alpha)$ and $DC_{s_0}^{(sim)}(\alpha)$ represent the prediction band and domain coverage for the $sim^{th}$ simulation, $sim=1,\ldots,S$;
in the simulation study that follows $S=100$. In practice, the functional coverage depends somehow on the discretization that one makes of the corresponding curve $Y_{s_0}(t)$, in the sense that the finer the discretizing grid, the more difficult it is for the whole curve to be contained in the band, and indeed the functional coverage varies slightly depending on the number of points chosen to represent the curve. That said, we opt for a fine discretization (101 points) but allow for a small tolerance $\varsigma$ in the functional coverage, in the sense that rather than requiring the whole true curve to be contained in the band, we require at least $100(1-\varsigma)\%$ of the true curve to be contained in the band. This modified version of the functional coverage with tolerance $\varsigma$ can be evaluated as
\begin{equation}\label{FC}
	FC_{s_0}^{\varsigma}(\alpha)=\frac{1}{S}\sum_{s=1}^S\mathbb{I}\left\{DC_{s_0}^{(sim)}(\alpha)\geq 1-\varsigma\right\}.
\end{equation}
Obviously, $FC_{s_0}^{\varsigma}(\alpha)$ coincides with $FC_{s_0}(\alpha)$ when $\varsigma=0$.

\section{Simulation Study}
\label{SIMU}

The performance of the bootstrapping method proposed in Section~\ref{BOOT} is evaluated here through a simulation study. Our aim is to analyse the impact of the spatial structure of the functional residual random field, by means of the covariance function parameters (scale and range), as well as that of the ordering technique chosen to derive the uncertainty bands when increasing the number of sites.

Data were simulated using cubic B-splines on a spatial irregular grid ($n$ locations) on $D=[0,2]\times[0,3]$  and curve domain $T=[0,1]$. We used a B-spline basis on $T$ with 10 basis functions. The residual functional random field was built as $e_s(t)=\sum_{j=1}^{10}\xi_j(s)B_j(t)$,
where $B_j(t)$ is the $j^{th}$ basis function evaluated at $t \in T$ (i.e. a curve) and $\{\xi_j(s),s \in D\}$ are the spatially correlated spline coefficients. These were generated independently for each $j=1,\ldots,10$ using the same exponential covariance function with range and scale parameters  $\phi \in (0.5,1,1.5)$ and $\sigma^2 \in (0.25,0.50,0.75)$ respectively, resulting in 9 different scenarios.
As already mentioned in Section~\ref{FKED} and shown in the appendix, the trace-variogram of $\{e_s(\cdot) ,s \in D\}$ turns out to be proportional to the variogram of the spline coefficients $\{\xi_j(s),s \in D\}$. 
The drift was obtained as
\[
	m_s(t)=\alpha(t) + \beta_1(t)lon + \beta_2(t)lat
\]
where $lon$ and $lat$ are the spatial coordinates, $\alpha(t)$ is a functional intercept and $\beta_1(t),\beta_2(t)$ are functional coefficients. 
The functional coefficients $\alpha(t)$, $\beta_1(t)$ and $\beta_2(t)$ can be expressed in terms of the same B-spline basis with scalar spline coefficients (common for all sites) that are  drawn 
from normal distributions as follows: for $\alpha(t)$ we draw 10 splines coefficients from $N_{10}(\mathbf{1}, 0.05\, I_{10\times 10})$; for $\beta_1(t)$ and $\beta_2(t)$ we draw 10 splines coefficients from $N_{10}(\bm{\vartheta}, 0.05 \, I_{10\times 10})$ with $\bm{\vartheta} =(0.2,0.2,0.4,0.4,0.6,0.6,0.8,0.8,1,1)^T$, where $I_{10\times 10}$ represents the identity matrix with dimension $10\times 10$.
Finally, simulated observations were built as
\[
	Y_{s}^{sim}(t) = m_s(t) + e_s(t) + \eta_s(t)
\]
where $\bm{\eta}(t)=\{\eta_{s_1}(t), \ldots, \eta_{s_n}(t) \} \sim N_{n}(\mathbf{0}, 0.09\, I_{n\times n})$ is a vector of random errors for each fixed $t \in [0,1]$; in practice we consider $M=$101 equally spaced points in $[0,1]$.

\begin{figure}[h!]
\begin{center}
\includegraphics[scale=0.6]{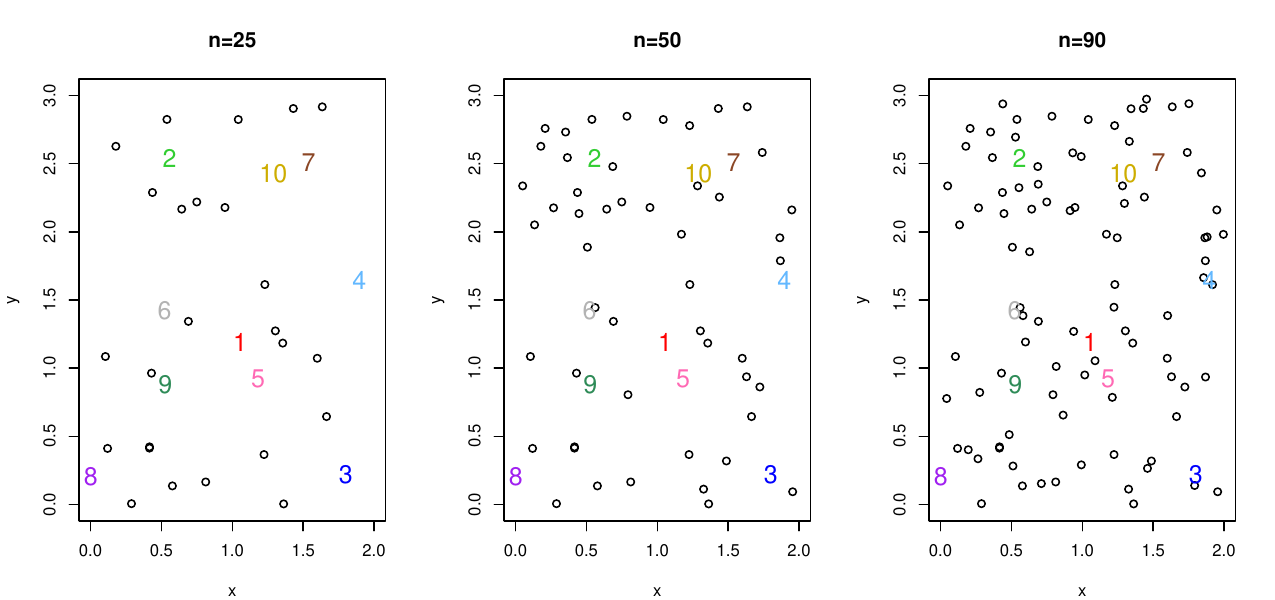} 
\end{center}
\caption{Locations of the 25, 50 and 90 sites used for model fitting. Validation sites numbered 1 to 10.}\label{simulation_sites}
\end{figure}

\begin{figure}[h!]
\begin{center}
\includegraphics[scale=0.68]{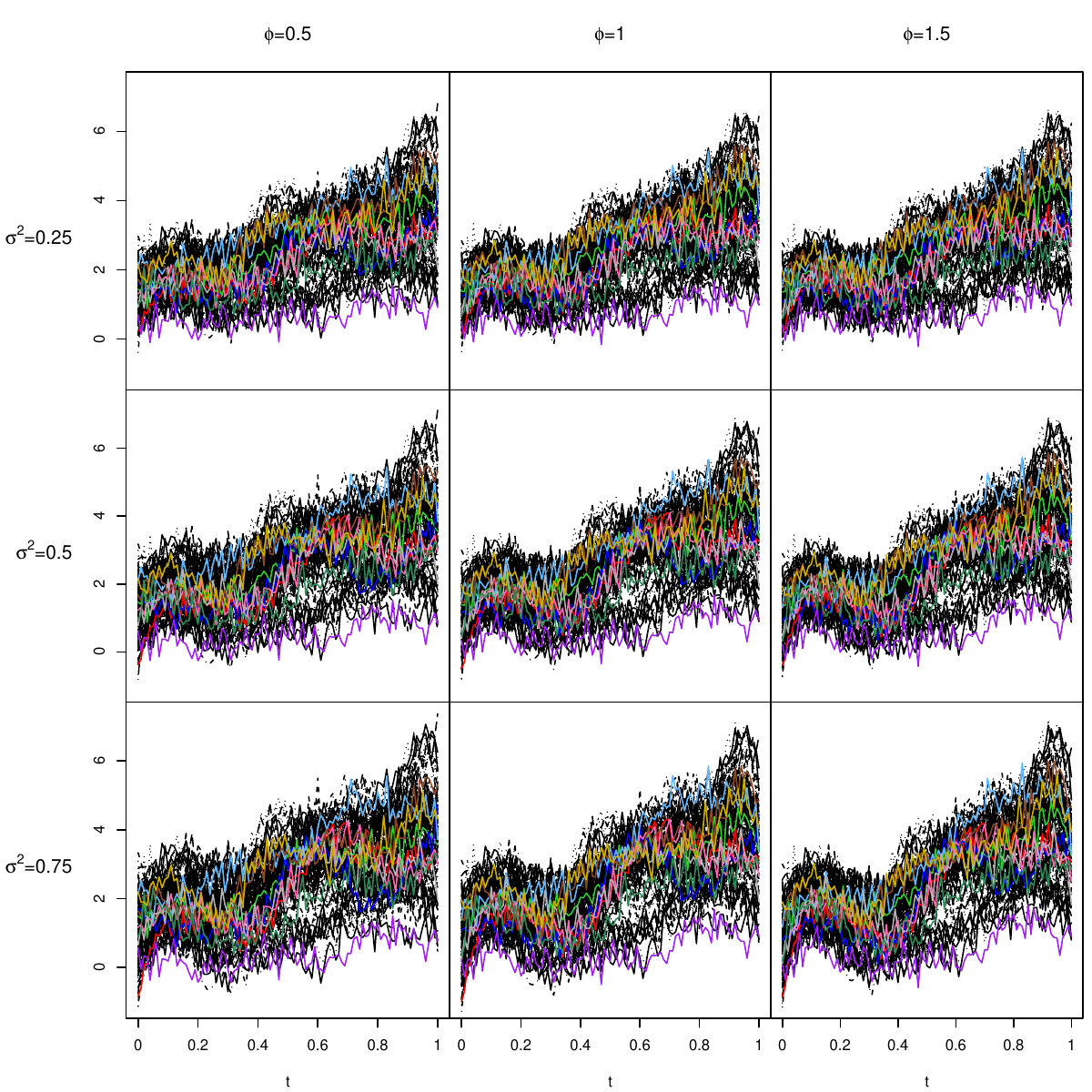}
\caption{Simulated data (n=90) with 10 validation stations curves in color, for each simulation scenario.}\label{simulation_data_n100}
\end{center}
\end{figure}

For each simulation scenario, we generated functional data at $n=25,50$ and 90 nested locations. Additionally, data were generated at 10 more sites (always the same for all three sample sizes) used as validation stations. 
Note that these validation stations, numbered 1 to 10 throughout the paper, represent a number of different situations that could be found in real applications, ranging from isolated locations with no information nearby (e.g. number 3) to locations situated very close to sites where data are available (e.g. number 9).
The locations can be seen in Figure~\ref{simulation_sites}, while the simulated data can be seen in Figure~\ref{simulation_data_n100} for $n=90$ (note that the cases $n=25$ and $n=50$ are just subsets of this).

\begin{figure}[t]
\begin{center}
\includegraphics[scale=0.5]{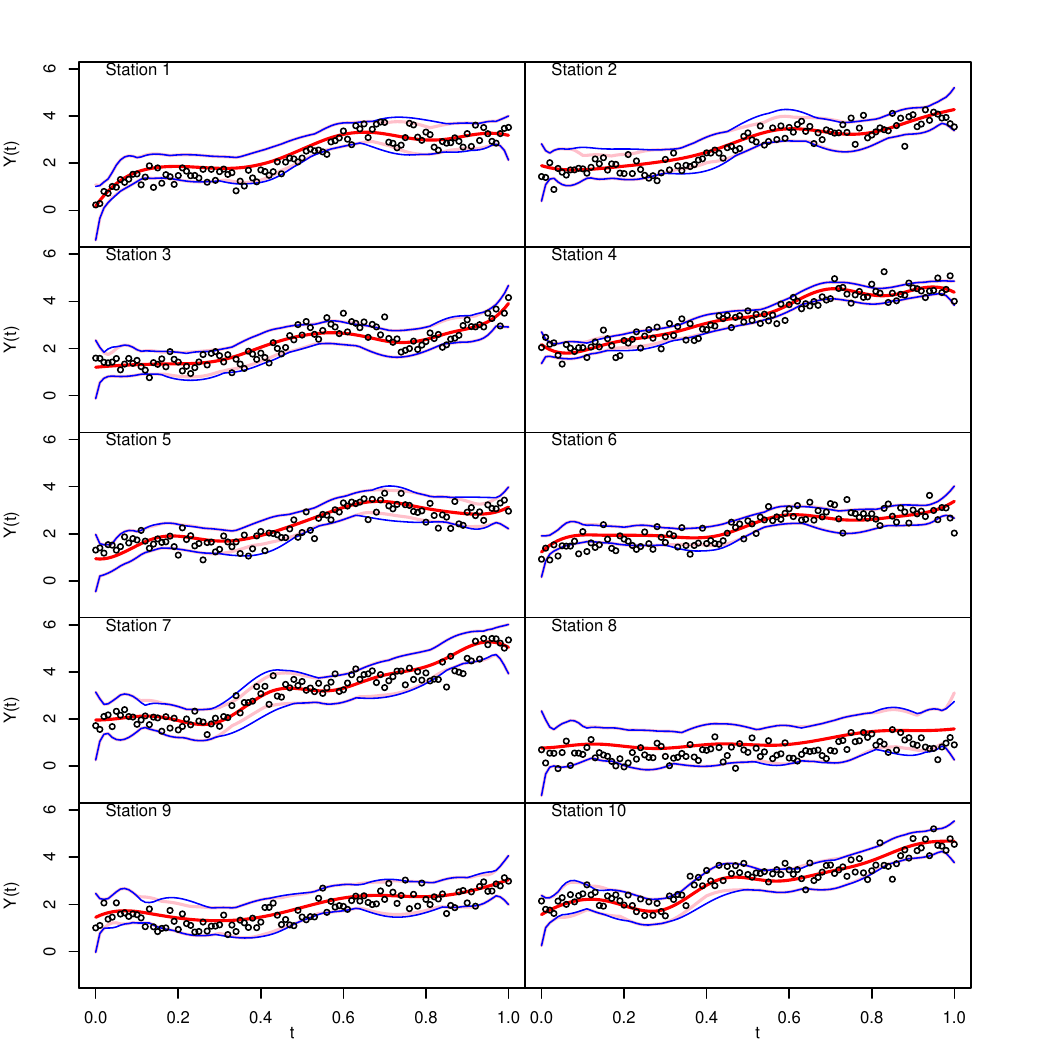}
\caption{Original data (black dots), FKED predicted curve (solid red line), 95\%  prediction band based on $L^2$ distance (pink) and on \emph{MBD} (blue) for n=90, $\sigma^2=0.25$, $\phi=0.5$.}\label{sim1_n100_fitCI}
\end{center}
\end{figure}

For each of the three sample sizes, and each simulation scenario, the FKED model presented in Section~\ref{FKED} was applied to the corresponding data set to predict curves at the 10 validation sites. In particular, a drift term depending on longitude and latitude was considered and ordinary kriging was used to obtain the predicted residuals. In practice, the variogram model is chosen automatically among exponential, gaussian and spherical based on minimum SSE. Computational times ranged from 2 seconds (n=25) up to 16 seconds (n=90).
The resulting predicted curves, along with the observed data, can be seen in Figure~\ref{sim1_n100_fitCI} in the case of $n=90$ and range and scale parameters $\phi=0.5$ and $\sigma^2$=0.25, respectively. 
The estimated range and scale parameters $\hat{\phi}$ and $\hat{\sigma}^2$ for all 9 simulation scenarios are in agreement with what one would expect given the values set for $\phi$ and $\sigma^2$ in the simulation design and the relationship between the trace-variogram and the variogram described in the appendix.
It appears that there is good accordance between simulated and predicted observations. Similar figures for the remaining cases are available upon request  as supplementary material.
Following the algorithm illustrated in Section~\ref{BOOT}, a bootstrap sample of size $B=500$ was obtained for each validation station.  
These 500 curves were ordered using both distance and band depth, where the latter was calculated using the modified version \emph{MBD}. The resulting 95\% prediction bands are shown in Figure~\ref{sim1_n100_fitCI}. Overall, the two uncertainty measures provide very similar prediction bands.
 
\begin{figure}
\begin{center}
\includegraphics[scale=0.55]{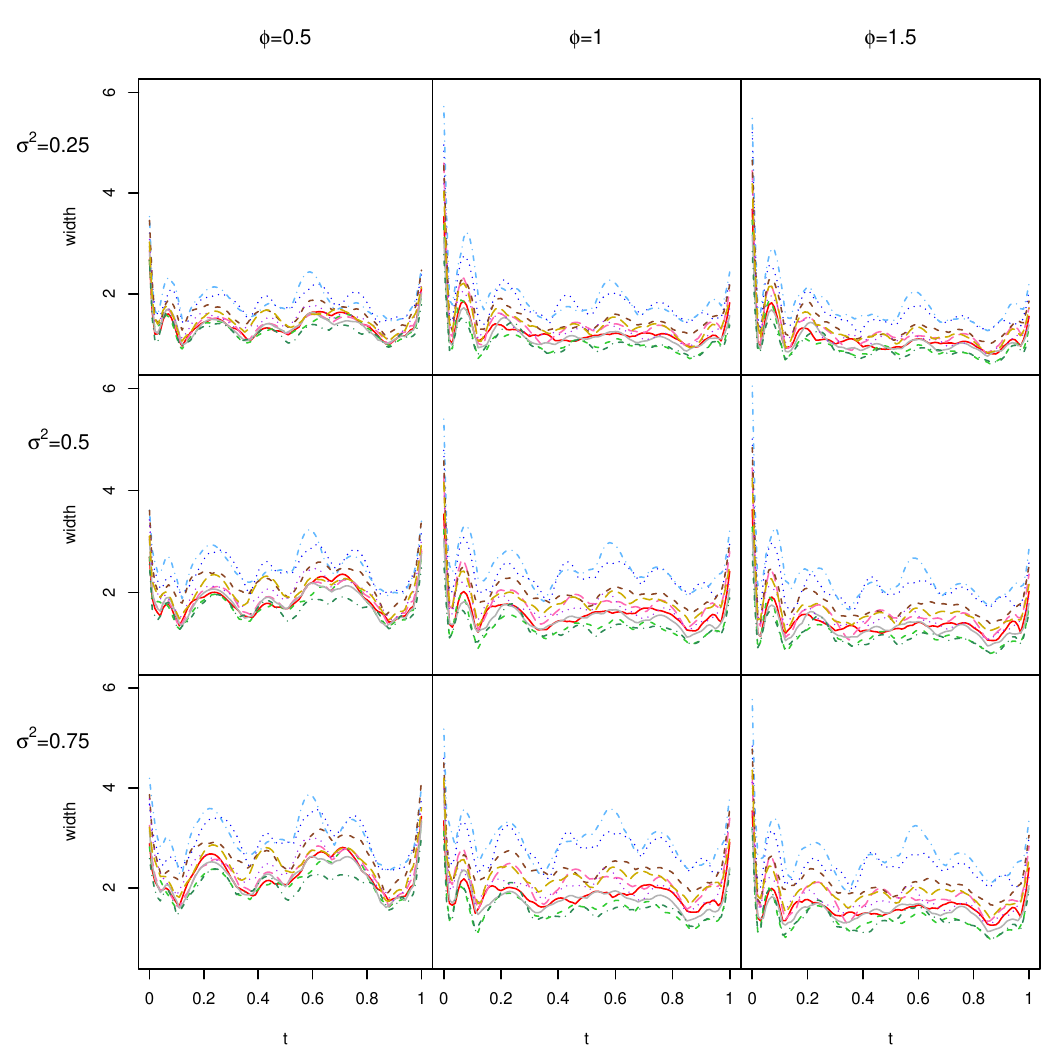}
\includegraphics[scale=0.55]{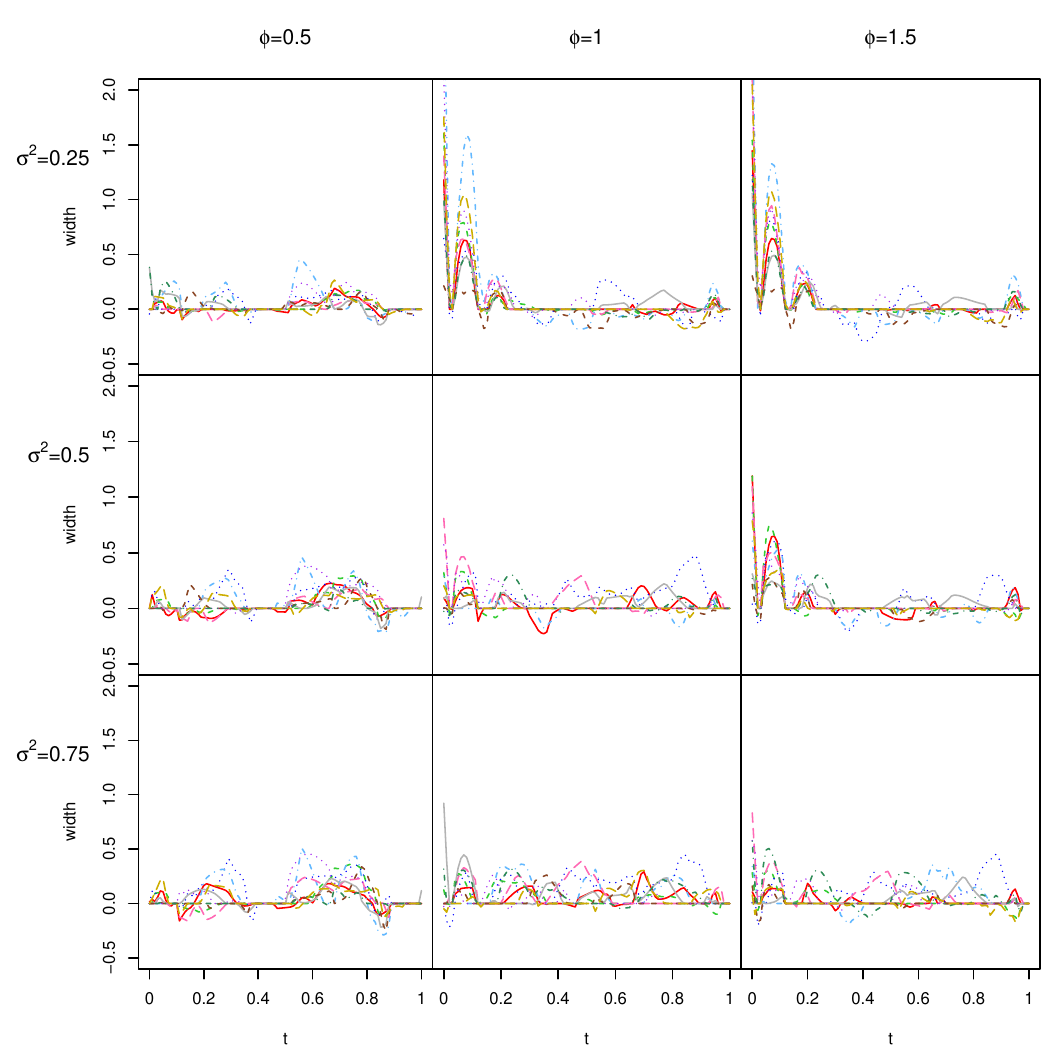}
\end{center}
\caption{Prediction band width according to band depth (top) and width (depth-distance) difference (bottom) for 10 validation stations and different simulation scenarios when $n=25$.}\label{simulation_width_n35}
\end{figure}

\begin{figure}[h!]
\begin{center}
\includegraphics[scale=0.68]{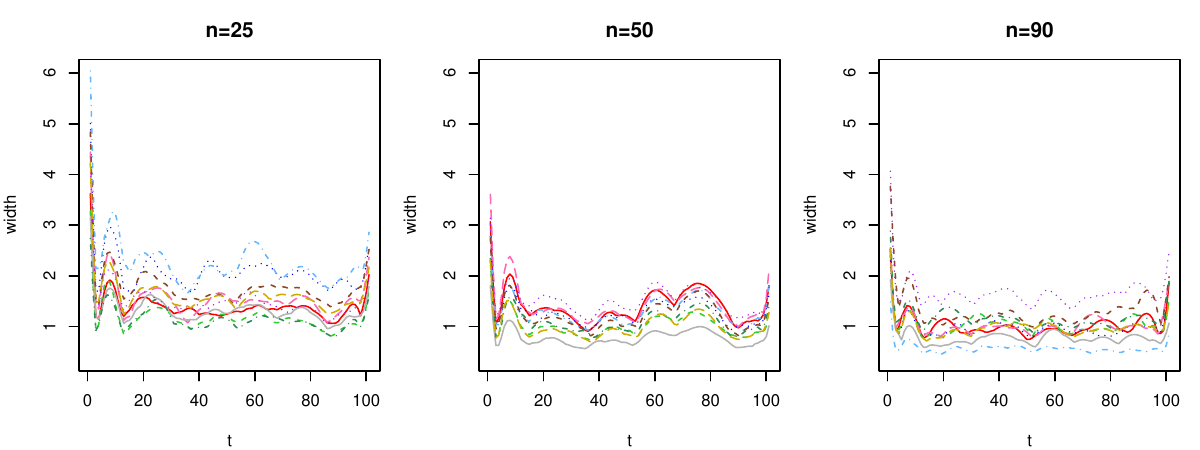}
\caption{Prediction band width, when using band depth for ordering curves, for a fixed simulation scenario ($\sigma^2=0.5$ and $\phi=1.5$) and all three samples sizes.}\label{sim6_width_depth_alln}
\end{center}
\end{figure}

In Figure~\ref{simulation_width_n35} (top), we show the depth based band width corresponding to a sample size of $n=25$ and all simulation scenarios, while Figure~\ref{simulation_width_n35} (bottom) summarizes the difference in width when using band depth and distance for the same sample size. The remaining figures (for $n=50,90$) are not shown here due to space limitations but are available upon request as supplementary material.
Greater values of the scale parameter $\sigma^2$ lead to wider prediction bands, while for a fixed value of the scale parameter, the width of the band decreases slightly with increasing range $\phi$.
When comparing band depth and distance (Figure~\ref{simulation_width_n35} (bottom)) it is interesting to see that the depth based interval is predominantly wider than the distance based one, regardless of the value of $\sigma^2$ and $\phi$. 
Figure~\ref{sim6_width_depth_alln} shows the depth based prediction band width for a fixed simulation scenario ($\sigma^2=0.5$ and $\phi=1.5$) and all three samples sizes; here it can be seen that as $n$ increases, the width of the interval decreases and it becomes slightly more stable on $T$.

\begin{figure}[t]
\begin{center}
\includegraphics[scale=0.68]{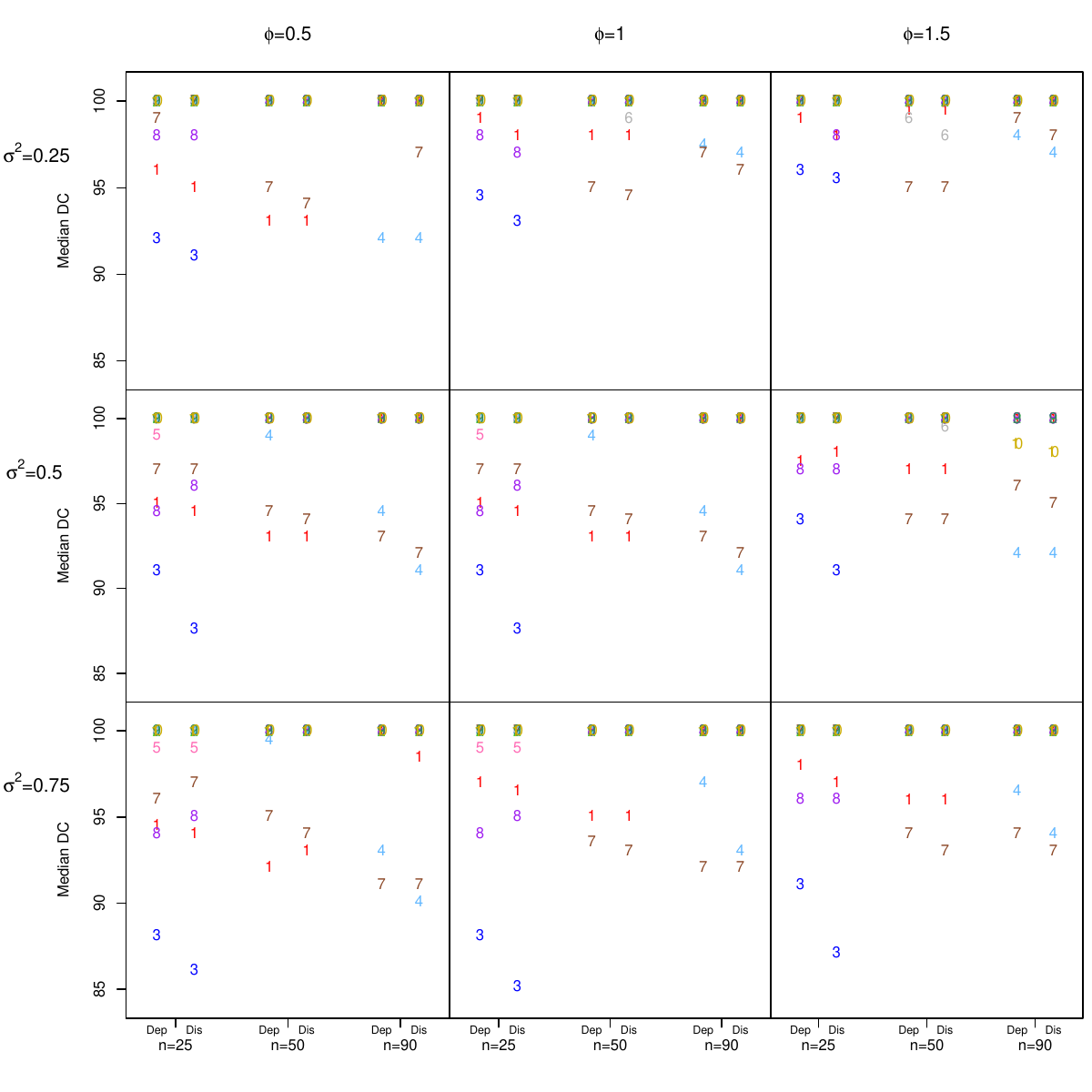}
\caption{Median domain coverage (over $S=100$ simulations) at each validation station (numbered 1 to 10) for $n$=25, 50 and 90 and all 9 simulation scenarios.}\label{simulation_coverage_boxplot}
\end{center}
\end{figure}

On the other hand, the median domain coverage (see Eq.~(\ref{DC})) over $S=100$ simulations is summarized in Figure~\ref{simulation_coverage_boxplot} for all simulation scenarios and sample sizes. Looking at the empirical distribution of $DC_{s_0}(\alpha)$ over $S=100$ simulations the maximum is always 100\%, whereas the minimum varies from 78.2\% to 100\%. This coverage is linked to the width of the interval (the larger the width the larger the coverage) but also to the goodness of FKED prediction because the uncertainty band is built around the predicted curve: if the prediction is far from the original data, the corresponding coverage will be poor, and viceversa. 
We can see in Figure~\ref{simulation_coverage_boxplot} that median domain coverage improves as $n$ increases and it is slightly better for small values of $\sigma^2$. Overall, median coverage ranges from 85.2\% to 100\% and is highly dependent on validation site. In particular, it can be seen that when $n=90$ all locations apart from validation stations numbered 4 and 7 have a median domain coverage greater than 95\%.
Station 3 seems to perform particularly badly when $n=25$ but this is explained by the fact that it is located on the border of the spatial region considered (see Figure~\ref{simulation_sites}(left)) and has no neighbouring sites.

\begin{figure}
\begin{center}
\includegraphics[scale=0.5]{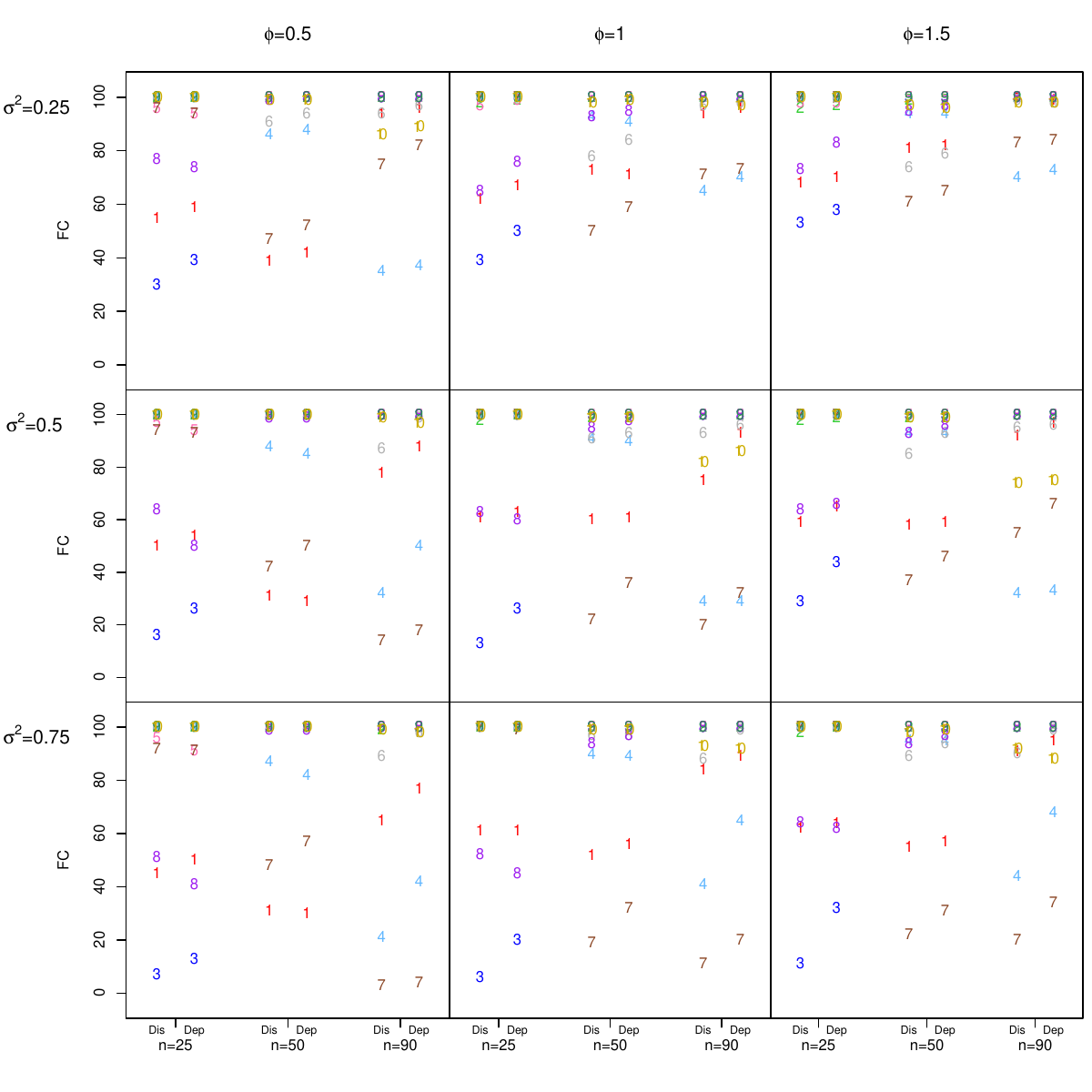}
\includegraphics[scale=0.5]{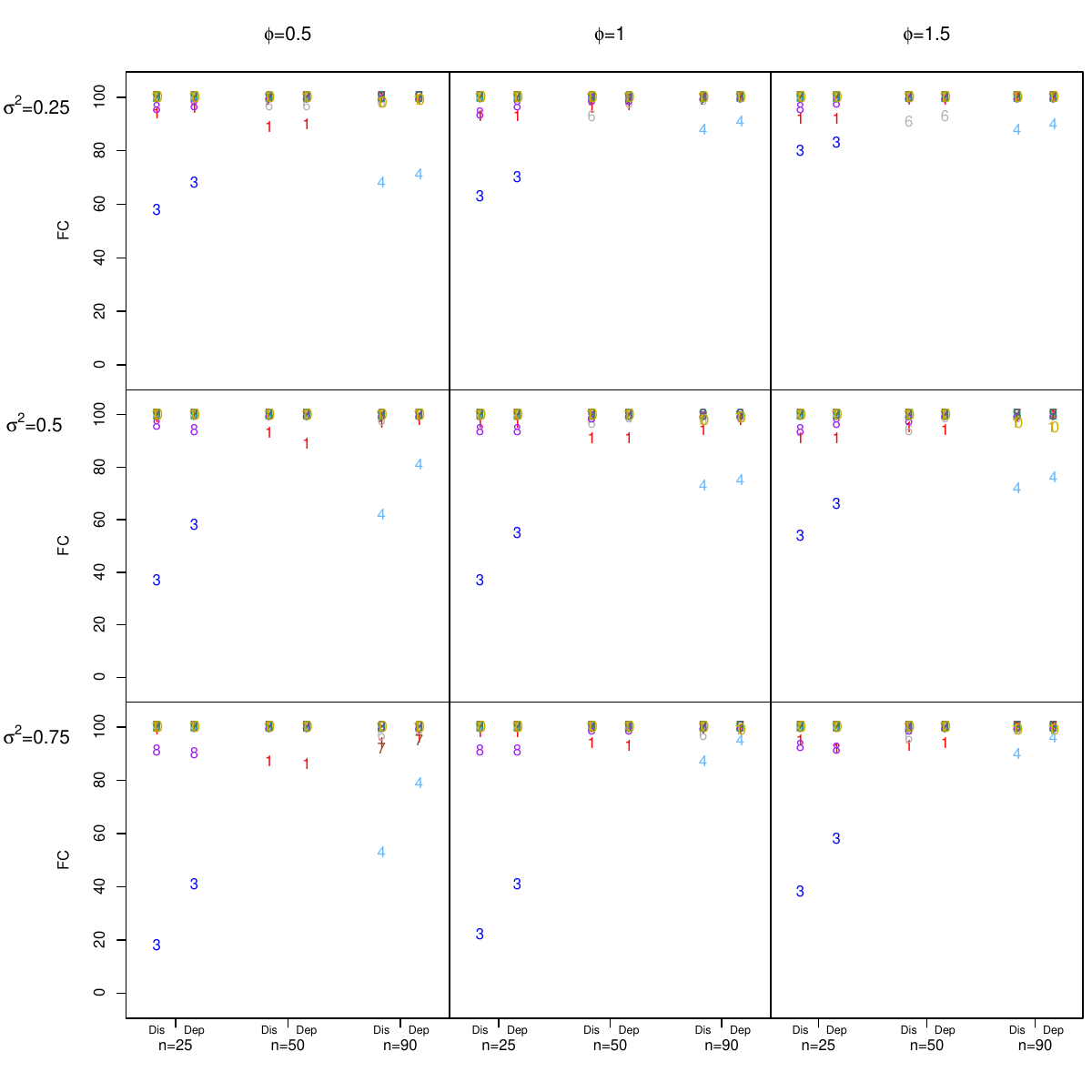}
\end{center}
\caption{Functional coverage $FC^{\varsigma}(0.05)$ at each validation station (numbered 1 to 10) for $n$=25, 50 and 90 and all 9 simulation scenarios with tolerance $\varsigma=0.05$ (top) and $\varsigma=0.10$ (bottom).}
\label{simulation_func_coverage}
\end{figure}

Following the earlier discussion at the end of Section~\ref{BOOT}, we allow for a small tolerance when calculating functional coverage (see Eq.~(\ref{FC})), and this is shown in Figure~\ref{simulation_func_coverage} for $\varsigma=~0.05$ and $\varsigma=~0.10$. It can be seen that $FC^{\varsigma}_{s_0}(0.05)$ decreases as $\sigma^2$ increases but improves with increasing values of $\phi$ (i.e. when the dependence structure becomes stronger) and varies greatly with validation site. Once again, when $\varsigma=0.05$ (Figure~ \ref{simulation_func_coverage} top) sites 4 and 7 show a lower functional coverage for $n=90$; if we disregard these two, $FC^{0.05}_{s_0}(0.05)$ for the remaining stations is very close to 100\% in the best scenario ($\sigma^2=0.25$, $\phi=1.5$).
Given the empirical distribution of domain coverage discussed above, it is intuitive to see that we cannot expect a functional coverage around the nominal level 95\% for all validation stations. For example, the minimum and median domain coverage for station 4 when $n=90$, $\sigma^2=0.75$, $\phi=0.25$ are 82.2\% and 90.1\% respectively, while the $5^{th}$  percentile is 84.2\%; this means that if we set the tolerance ${\varsigma}=0.05$ the nominal level 95\% cannot be reached. 
If instead we allow for a tolerance ${\varsigma}=0.10$ (see Section~\ref{sec:conc} for a discussion), the functional coverage improves considerably as shown in Figure~\ref{simulation_func_coverage}(bottom).
Moreover, if rather than discretizing the curve to 101 points we use $M=50$, the results on coverage improve; for example, median domain coverage ranges from 90\% to 100\%.

\section{Real data analysis}

\subsection{Canadian temperature data}
\begin{figure}[h!]
\begin{center}
\includegraphics[scale=0.68]{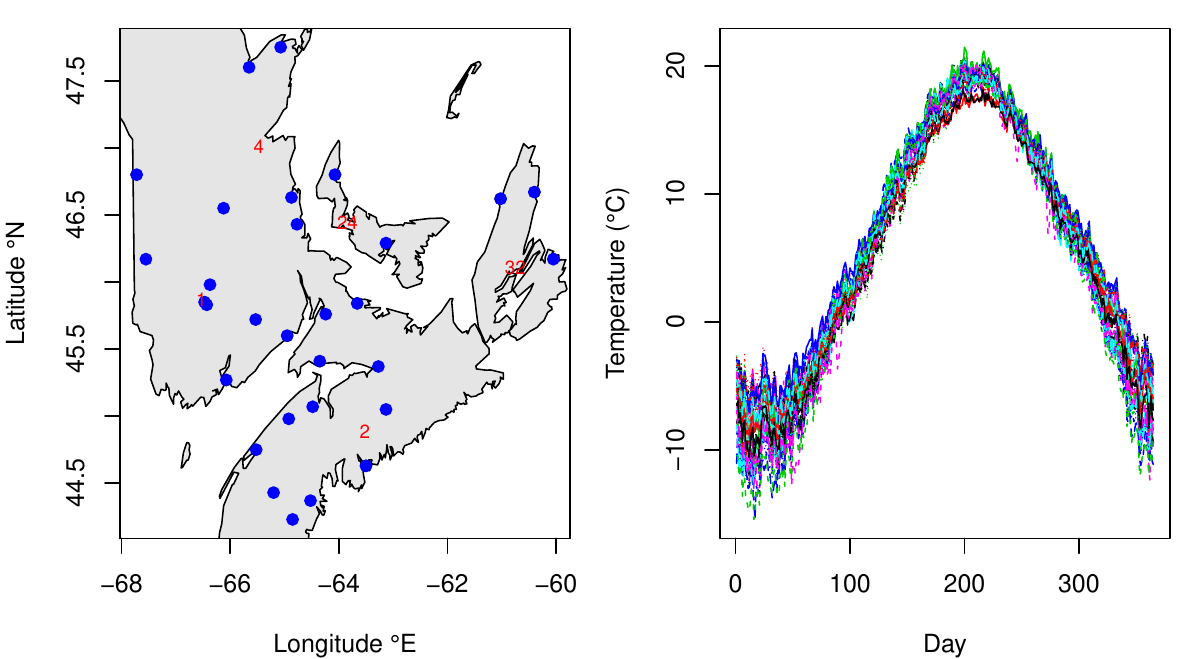}
\caption{Locations of the 35 meteorological stations in Canada's Maritimes Provinces area (left, validation stations numbered in red) and temperature curves (right, raw data)}\label{canada_map_data_val_numbered}
\end{center}
\end{figure}

\begin{figure}[h!]
\begin{center}
\includegraphics[scale=0.4]{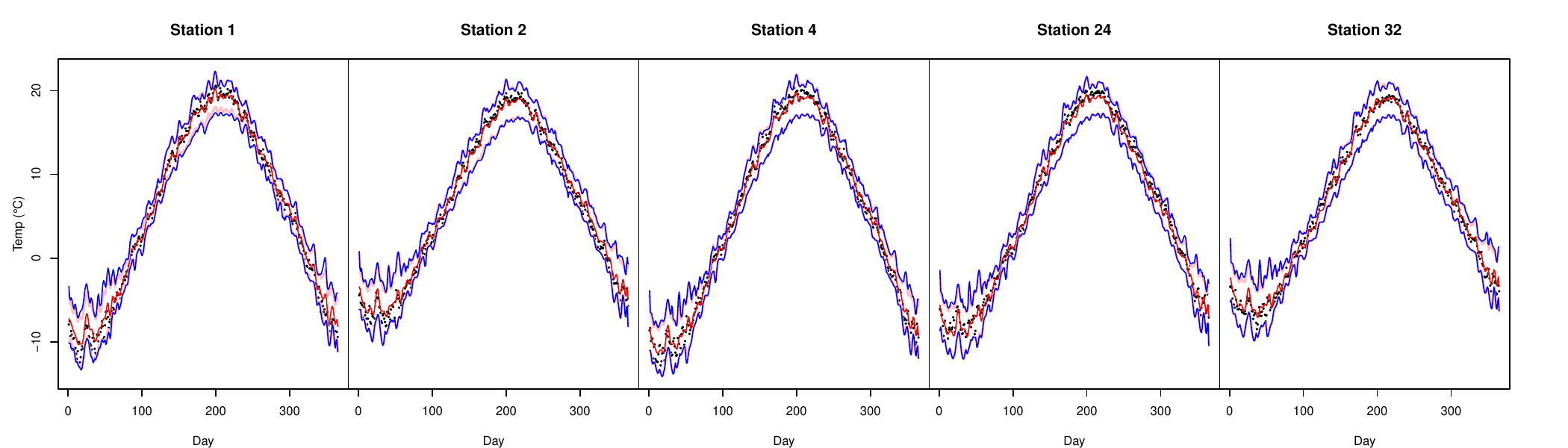}\\
\caption{Original temperature data (black dots), FKED predicted curve (red line), 95\% prediction band based on $L^2$ distance (pink) and on \emph{MBD} (blue) for validation stations}\label{CI_canadatemp_k120_smo0_05}
\end{center}
\end{figure}

As first case study, we have chosen the well known data set of Canadian temperature. This has been repeatedly used in the functional data literature (see, for example, \cite{Menafoglio2013, Giraldo2010, FDAbook,  FAMM}). The data set consists of daily annual mean temperature collected at 35 meteorological stations in Canada's Maritimes Provinces: Nova Scotia, New Brunswick and Prince Edward Island, over the period 1960 to 1994. Note that the data set used here (available in the \texttt{geofd R} package; \cite{geofd}), which is the same as in \cite{Menafoglio2013, Giraldo2010}, covers a smaller geographical area than that in \cite{FDAbook}. While other papers have been devoted to prediction of these temperature curves, our objective is to provide uncertainty bands for the predicted curves. Therefore, discussing the fitted model is out of the scope of this paper. For this aim, we selected five stations at random to use as validation stations for which we will provide prediction bands according to our proposal in Section~\ref{BOOT}. These are marked in red in Figure~\ref{canada_map_data_val_numbered} (left). Data were converted to functional observations through smoothing by using penalized cubic B-splines with 120 basis functions and penalty parameter equal to zero. These values were chosen using functional cross-validation. 
The FKED model, with longitude and latitude as covariates, was then fitted to the remaining 30 stations and predicted temperature curves were obtained using ordinary functional kriging with an exponential variogram model for the 5 validation stations. From the empirical trace-variogram, there was no evidence of a discontinuity at the origin and hence we fixed the nugget equal to zero. 
For each of the validation sites, a bootstrap sample of size $B=1000$ was obtained. Band depth was calculated using the modified version \emph{MBD}. The resulting 95\% prediction bands are shown in Figure~\ref{CI_canadatemp_k120_smo0_05}. 
The uncertainty bands are fairly narrow, as expected when observing the small variability among curves in Figure~\ref{canada_map_data_val_numbered} (right) but they become slightly wider in winter. Overall, the two uncertainty measures seem to agree well, although in some cases the distance based prediction band appears to be slightly narrower than the depth based one. 
Domain coverage percentages for all 5 validation sites range from 98.9\% to 100\%.

\subsection{Air pollution data}
\begin{figure}
\begin{center}
\includegraphics[scale=0.4]{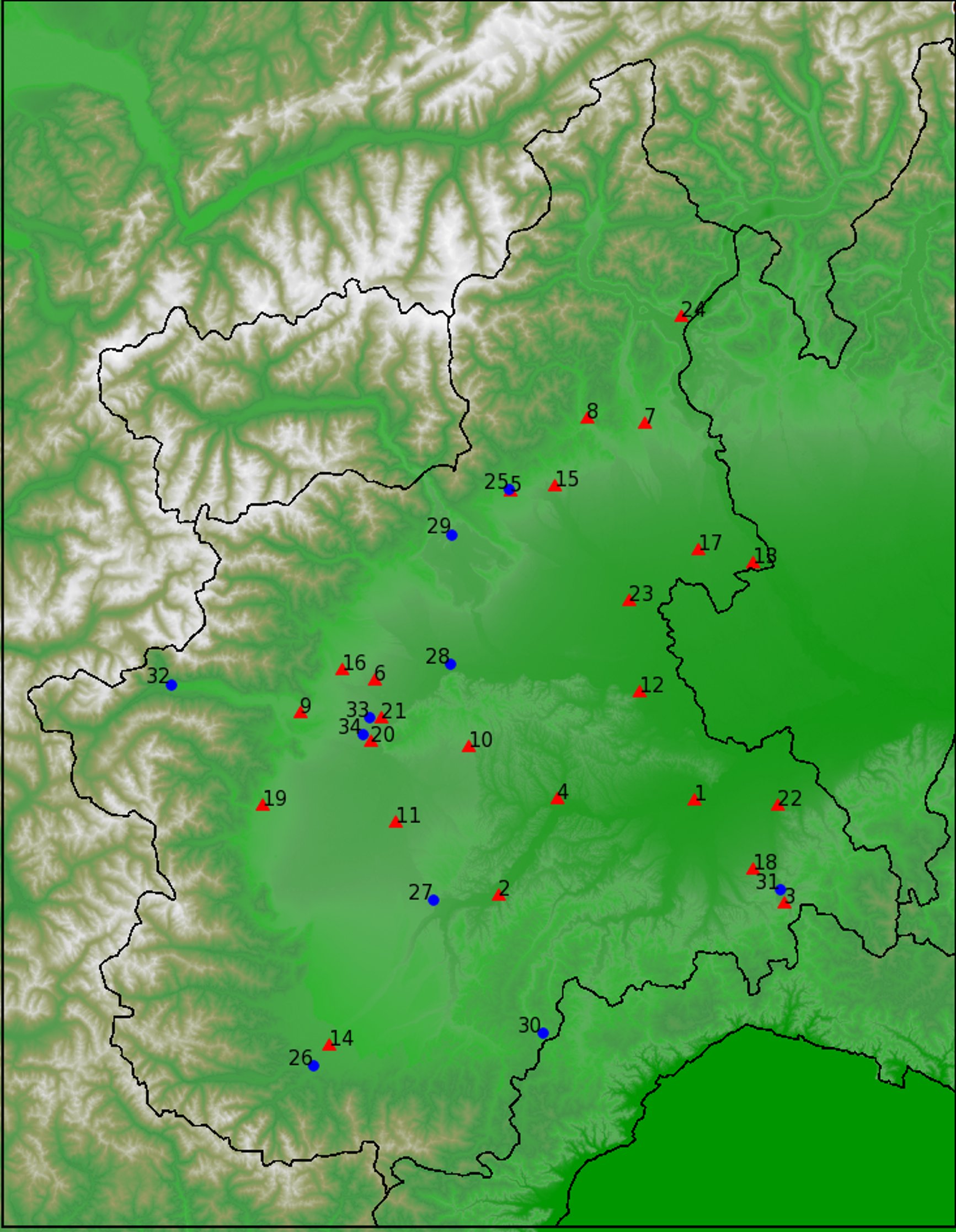}
\caption{Locations of the 24 PM$_{10}$ monitoring sites (red triangles) and 10
validation stations (blue dots).}\label{mappa_stazioni}
\end{center}
\end{figure}

The second case study considered consists of daily PM$_{10}$ concentrations (in $\mu g/m^3$) measured in 24 sites (red triangles in Figure~\ref{mappa_stazioni}) from October 2005 to March 2006 by the monitoring network of Piemonte region (Italy). Measurements were also available at 10 additional locations (blue dots in Figure~\ref{mappa_stazioni}) that are used as validation stations. Apart from geographical information, i.e. longitude, latitude and altitude of each station, which were considered as scalar covariates, information was available on daily maximum mixing height, daily total precipitation, daily mean wind speed, daily mean temperature and daily emission rates of primary aerosols, which were taken as functional covariates. These are available as a result of a nested system of  deterministic computer-based models implemented by the environmental agency ARPA Piemonte \citep{Finardi2008}. This data set had already been analyzed in \cite{Cameletti2011}. For further details on the model the reader is referred to \cite{IgnaccoloSERRA}. A log transformation was used on the response variable to achieve normality and stabilize within-station variability.
Prior to modelling, data (both response and functional covariates) were smoothed by means of cubic B-splines with 146 basis functions and penalty parameter equal to 0. These values were chosen using functional cross-validation \citep{IgnaccoloSERRA}.

\begin{figure}[h!]
\begin{center}
\includegraphics[scale=0.85]{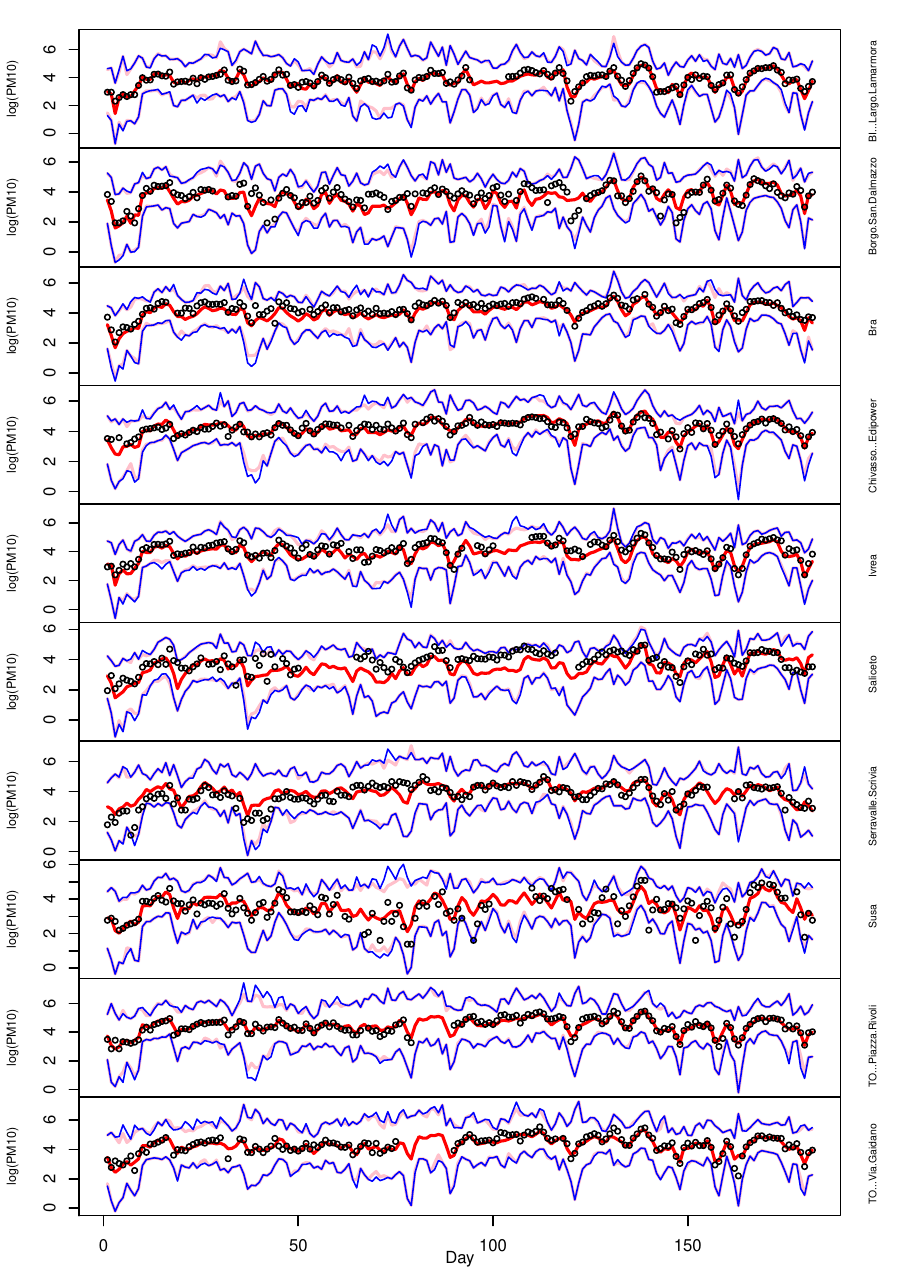}\\
\caption{Original PM$_{10}$ data (black dots), FKED predicted curve (red line), 95\% prediction band based on $L^2$ distance (pink) and on \emph{MBD} (blue) for validation stations.}\label{CI_pmvalidation}
\end{center}
\end{figure}

The functional kriging with external drift model described in Section~\ref{FKED} was applied to the air pollution data, including the (standardized) scalar and functional covariates  mentioned above, to obtain prediction curves (via ordinary kriging for functional data with an exponential variogram model and zero nugget) at the 10 validation sites.
To obtain 95\% prediction bands for each of the predicted curves, a bootstrap sample of size 1000 was obtained for each site following the algorithm proposed in Section~\ref{BOOT}. 
We obtained prediction bands according to both the modified band depth (with $k=2$) and distance induced order. Prediction bands for the ten validation sites are shown in Figure~\ref{CI_pmvalidation}.
Overall, the two uncertainty measures seem to agree well, although in some cases the depth based band appears to be slightly wider than the distance based one. The domain coverage varies from 97.3\% to 100\%.

\section{Discussion}
\label{sec:conc} 
The functional kriging with external drift model allows to predict a whole curve - regardless of the domain of the functional observations - taking into account exogenous covariates and the underlying spatial dependence. Nevertheless, uncertainty evaluation in the functional kriging context has hardly ever been addressed in the literature. 
The classic functional kriging variance provides a unique value over the whole domain $T$, while our aim is to provide an uncertainty measure whose value may change along the predicted curve. Given the lack of an analytic expression of a domain-varying kriging variance for a curve, we propose a semiparametric bootstrap approach that not only allows the uncertainty to change over the domain $T$ but also takes into account the uncertainty due to drift estimation. 
The contrast considered to build the uncertainty band, defined as the difference between the spatial prediction and the unknown value, could be thought of as a sort of pivotal quantity, the distribution of which is unknown. The use of a pivotal quantity is recalled to improve the performance of the bootstrap in the parametric framework \citep{Canty2006}.

We considered two different techniques proposed in the literature for ordering the bootstrapped curves, namely functional depth and $L^2$ distance, but we did not find great differences in the results for either the simulations or the two case studies. To be conservative, one could argue that it is safer to use the band depth as this criterion gives larger bands on average. 

Overall, using the known data at validation sites in both simulations and real case studies, it can be concluded that the proposed method is a valid approach in evaluating uncertainty for a predicted curve.  
In particular, the simulation study shows that the width of the bands depends on the scale and range parameters of the covariance structure, at least for the exponential case, and becomes more stable over the domain $T$ with increasing sample size. 

In terms of functional coverage, we can conclude that once we allow for a small tolerance the method performs generally well. If for a nominal domain coverage we are willing to accept an effective coverage of 90\%, then it seems reasonable to assume a tolerance $\varsigma=0.10$, in which case results are satisfactory.
Note that, in practice, when the method is applied to a real dataset, the functional coverage cannot be calculated and assessment would have to be done based solely on domain coverage, whose median in the simulation study was not far from the nominal level of 95\%.

Data in the two case studies considered in this paper were regularly spaced over the domain $T$; however, that is not always the case (see e.g. atmospheric profiles in \cite{IgnaccoloCollocation}). For irregularly spaced data, curves could be aligned at the initial smoothing step before fitting the functional kriging with external drift model, as it is straightforward to evaluate the curves for every $t \in T$ in a common grid for all curves.

The algorithm proposed for uncertainty evaluation of spatially correlated curves is computationally feasible (running times with $B=1000$: 4 hours for the PM$_{10}$ case study and 2.35 hours for the Canadian temperature case study using an Intel Core i7-4770 CPU 3.40GHz 16GB RAM) and applies to a wide range of practical situations, as it can be used regardless of the complexity of the drift term, or even in the absence of the latter. 
Our proposal has been specified in the case of ordinary kriging where the weight coefficients are constant and the spatial structure is determined by means of the trace-variogram. 
We are aware that the simulation study is limited to the simple case in which the spatial dependence structure does not change w.r.t. $t \in T$, as assumed by the trace-covariogram. 
While it might be worth exploring more complex scenarios a full analysis of the air pollution data described in Section 5.2, \cite{IgnaccoloSERRA} showed that in practice the assumption of a spatial structure that is constant over time is reasonable, as the drift term in the model picks up most of the spatial variability over time. This suggests that in similar situations, where the available covariates are able to explain a large part of the time-space variability, ordinary residual kriging should be appropriate.

Nevertheless, it may be the case that a more complex kriging alternative is desiderable in order to let the weights vary with $t$ (for both the real data and the bootstrap samples). To be consistent with the way the kriging weights have been specified in the cases of continuous time-varying kriging and functional kriging total model, the underlying spatial structure (and hence the matrices $K$ in Section~\ref{FKED} and $\Sigma$ in Section~\ref{BOOT}) should not be determined by means of the trace-variogram. In fact, in \cite{Giraldo2011} and \cite{Giraldo2009} functional data are expressed as linear combinations of splines and a Linear Model of Coregionalization is used to estimate cross-correlations among the spline coefficients; this way a possible interaction between the curve domain and the space domain is taken into account. Consequently the matrices $K$ and $\Sigma$ would need to be adjusted accordingly but with the added burden of an increased computational load.

We believe that the method proposed in this paper is appropriate in the framework of functional data and is able to provide uncertainty bands for a predicted curve in an unmonitored site. Further, half the width of the resulting prediction band could be considered as an approximate margin of error. We think this will prove useful for monitoring purposes and policy assessment where the uncertainty should always accompany the related prediction.

\section{Acknowledgments}
This work was supported by ``Futuro in Ricerca'' 2012 Grant (project no. RBFR12URQJ, named StEPhI) provided by the Italian Ministry of Education, Universities and Research.
The authors are grateful to all the participants of the StEPhI workshops for the fruitful discussions. 

%
%
%
%
%

%

\bibliographystyle{Chicago}

\bibliography{mybib_JCGS}

\section*{Appendix: note on the trace-variogram}
Let us recall that the trace-variogram is defined, for a zero-mean weakly-stationary isotropic process, as 
\[
\upsilon(h)=\int_T \frac{1}{2}Var\left(\epsilon_{s_i}(t)-\epsilon_{s_j}(t)\right)dt
\]
where $h=||s_i-s_j||$ represents the Euclidean distante between locations $s_i$ and $s_j$.\\
By assuming a finite development for the functions $\epsilon(t)$ such that $\epsilon_{s_i}(t)=\sum_{l=1}^{Nb} \xi_l(s) B_l(t)$ where $Nb$ is the number of considered basis functions and $B_l(t)$ is the $l$-th basis function evaluated at $t \in T$, we can write
\[
Cov\left(\epsilon_{s_i}(t),\epsilon_{s_j}(v)\right)=\sum_{l=1}^{Nb} Cov\left(\xi_l(s_i),\xi_l(s_j)\right) B_l(t)B_l(v)\\
= C(s_i, s_j) \sum_{l=1}^{Nb} B_l(t)B_l(v) = C(s_i, s_j)\tau(t,v)
\] 
where $C(s_i, s_j) =  Cov\left(\xi_l(s_i),\xi_l(s_j)\right)$ is assumed identical for all $l$ 
and $\tau(t,v)=\sum_{l=1}^{Nb} B_l(t)B_l(v)$. Thus we get a factorization with respect to the spatial domain $D$ and the curve domain $T$.\\
Moreover, for every $i$, $Var\left(\epsilon_{s_i}(t)\right) = C(s_i, s_i) \tau(t,t)$ and we can write

\begin{align*}
Var\left(\epsilon_{s_i}(t)-\epsilon_{s_j}(t)\right) &= Var\left(\epsilon_{s_i}(t)\right) + Var\left(\epsilon_{s_j}(t)\right) -2 Cov\left(\epsilon_{s_i}(t),\epsilon_{s_j}(t)\right)\\
&=C(s_i,s_i)\tau(t,t) + C(s_j,s_j)\tau(t,t) - 2 C(s_i,s_j)\tau(t,t).
\end{align*}
With the assumption of stationarity and isotropy for $\epsilon$ we obtain
\[
\upsilon(h)=  \left[ C(0) - C(h)\right] \int_T \tau(t,t)dt
\]
so that the trace-variogram is written as a product of a (classical) spatial variogram and a constant depending on the $Nb$ basis functions $B_l$.

\end{document}